\documentclass[12pt,titlepage]{utarticle}
\pdfoutput=1
\usepackage{amssymb,graphicx}
\usepackage{amsmath}
\usepackage{hyperref} 
\usepackage{cite}
\usepackage{upgreek}

\def\e{{\rm e}}

\def\eps{\epsilon}
\def\d{\partial}
\def\l{\left(}
\def\r{\right)}

\newcommand{\be}{\begin{equation}}
\newcommand{\ee}{\end{equation}}
\newcommand{\bea}{\begin{eqnarray}}
\newcommand{\eea}{\end{eqnarray}}
\newcommand{\bg}{\begin{gather}}
\newcommand{\eg}{\end{gather}}
\newcommand{\bseq}{\begin{subequations}}
\newcommand{\eseq}{\end{subequations}}

\begin{document}
\baselineskip=15.5pt
\title{Effective String Theory Revisited}

\author{Sergei Dubovsky
   \address{
      Center for Cosmology and Particle Physics, Department of Physics,\\
      New York University\\ 
      New York, NY, 10003, USA\\
   }
   , Raphael Flauger $^{1,}$ 
   \address{
      School of Natural Sciences,\\ 
      Institute for Advanced Study,\\
      Princeton, NJ 08540, USA\\
      {~}\\[.5cm]
      {\rm {\it e}-mail}\\[.1cm]
      \emailt{dubovsky@nyu.edu}\\
      \emailt{flauger@nyu.edu}\\
      \emailt{vg629@nyu.edu}\\
   } and Victor Gorbenko $^1$
}


\Abstract{We revisit the effective field theory of long relativistic strings such as confining flux tubes in QCD. We derive the Polchinski--Strominger interaction by a calculation in static gauge. This interaction implies that a non-critical string which initially oscillates in one direction gets excited in orthogonal directions as well. In static gauge no additional term in the effective action is needed to obtain this effect. It results from a one-loop calculation using the Nambu--Goto action. Non-linearly realized Lorentz symmetry is manifest at all stages in dimensional regularization. We also explain that independent of the number of dimensions non-covariant counterterms have to be added to the action in the commonly used zeta-function regularization.  
 }

\maketitle

\section{Introduction and Summary}
Long string-like objects are ubiquitous in field theory. Physical examples range from magnetic vortices in 
superconductors to (yet waiting to be observed) cosmic strings in grand unified models and to QCD flux tubes seen on the lattice \cite{Teper:2009uf,Kuti:2005xg}. 
Conceptually, these are very simple objects, at least from the effective field theory point of view.  In the absence of any additional light worldsheet degrees of freedom 
a long string  in a $D$-dimensional space-time is a system of $(D-2)$ two-dimensional Goldstone bosons $X^i$. 
Their purpose in life is to non-linearly realize the transverse spatial translations spontaneously broken by the presence of a string.  

The dynamics of an axially symmetric long string is then described by a generic action for $X$'s invariant under shift symmetries
\be
\label{shifts}
X^i\to X^i+x_0^i\;,
\ee
corresponding to non-linearly realized translations, and rotations 
\be
\label{rots}
X^i\to O^{i}_j X^j\;,
\ee
with $O\in SO(D-2)$.
As any effective field theory this action contains an infinite set of higher derivative operators suppressed by a length scale $\ell_s$. 
Physically, this scale corresponds to the width of the string, and the effective field theory description breaks down at distances shorter than $\ell_s$.

The system becomes more restricted when the underlying UV theory is Lorentz invariant, as for cosmic strings and QCD flux tubes.
 In this case the form of the action is further constrained because the full $D$-dimensional Poincar\'e group $ISO(D-1,1)$ must be non-linearly realized.
We will concentrate on this case in what follows.\footnote{Of course, there can be other interesting cases as well. For instance, for a vortex in an isotropic medium 
 the full rotation symmetry still imposes constraints on the low energy action. Galilean boosts are more subtle to implement in this case because those bring in also interactions with phonons of the medium.}
 Even though the number of broken generators has increased, the set of the Goldstone fields does not get enlarged. The reason is that space-time dependent translations include boosts as well.
 This subtlety in counting the Goldstone modes for space-time symmetries is well known. A relatively recent discussion can be found in \cite{Low:2001bw}.
 
 The task of constructing a general Lagrangian invariant under non-linearly realized Poincar\'e symmetry $ISO(D-1,1)$ is very similar to the problem of  constructing 
Lagrangians for Goldstone bosons corresponding to internal symmetries (with pions providing the primary example in particle physics). The latter was solved exhaustively by the Callan, Coleman, Wess and Zumino (CCWZ) construction \cite{Coleman:1969sm,Callan:1969sn}. Soon after CCWZ their recipe was generalized to spontaneously broken space-time symmetries \cite{Isham:1971dv,Volkov:1973vd}. For the case at hand it reduces to the following prescription.
Combine the fields $X^i$ and the world-sheet coordinates $\sigma^\alpha$ ($\alpha=1,2$) into a single object
\be
\label{Xmu}
X^\mu=(\sigma^\alpha,X^i(\sigma))\;,\;\;\mu=0,\dots, D-1\;.
\ee
These $X^\mu$'s are the coordinates of the embedding of the string worldsheet into the target space. Hence their transformation rules under the full Poincar\'e group $ISO(D-1,1)$ are simply those of the space-time coordinates. These are analogues of the sigma model $U$ field in the chiral pion Lagrangian.
The Lorentz invariant Lagrangian is then simply a sum of local geometric invariants constructed with the help of the embedding $X^\mu$,
\be
\label{coset_action}
S_{string}=-\int d^2\sigma\sqrt{-\det{h_{\alpha\beta}}}\l \ell_s^{-2}+{1\over\alpha_0}\l K_{\alpha\beta}^i\r^2+\dots\r
\ee
where $h_{\alpha\beta}$ is the induced metric on the world-sheet,
\be
\label{hind}
h_{\alpha\beta}=\d_\alpha X^\mu\d_\beta X_\mu
\ee
$K_{\alpha\beta}^i$ is the second fundamental form (the extrinsic curvature) of the world-sheet.
The first term in (\ref{coset_action}) is the Nambu--Goto (NG) action, the second one is the rigidity term introduced by Polyakov \cite{Polyakov:1986cs} and Kleinert
\cite{Kleinert:1986bk}, and dots stand for higher derivative geometric invariants.\footnote{Naively, at this order there are two additional operators, $({K^i}^\alpha_\alpha)^2$ and $R$. In two dimensions $R$ is a total derivative and the three operators are related by the Gauss-Codazzi equation so that in two dimensions only one of the extrinsic curvature squares has to be kept.} The tension of the string $\ell_s^{-2}$, the rigidity parameter $\alpha_0$, and the coefficients
in front of all other higher-derivative operators are free parameters of the low energy effective theory to be determined either from experiment (or from the lattice data for the QCD string), or from matching the effective theory to the microscopic theory in the UV (which can be done, for example, for  cosmic strings in weakly coupled models). 

Much of our discussion will deal with infinitely long strings because we are concerned with the form of the bulk action. IR effects such as finite size effects and boundary terms can be included at a later stage. (See {\em e.g.}~\cite{Aharony:2010db,Aharony:2010cx,Billo:2012da}.)

As expected, the action (\ref{coset_action}) is  invariant under the linearly realized $ISO(1,1)\times SO(D-2)$ symmetry, which is the unbroken subgroup of 
$ISO(D-1,1)$ in the presence of a long straight string. The  $ISO(1,1)$ factor  acts as a worldsheet Poincar\'e group, and $SO(D-2)$ acts as in (\ref{rots}). The remaining spatial translations act as in (\ref{shifts}), and the action of the remaining broken boosts and rotations $J^{\alpha i}$ following from the linear transformation law for $X^\mu$ is
\be
\label{non_boost}
\updelta^{\alpha i}_\eps X^j=-\epsilon( \delta^{ij}\sigma^\alpha+X^i\d^\alpha X^j)\;,
\ee
where $\epsilon$ is an infinitesimal  parameter of the boost/rotation.

Often as a starting point for formulating the string dynamics one chooses the manifestly covariant formalism, where all components of $X^\mu$ are considered as independent dynamical fields. Then the action (\ref{coset_action}) is invariant under an additional gauge symmetry, world sheet reparametrizations, and the formulation presented here arises as a result of gauge fixing defined by (\ref{Xmu}). The transformation  rule (\ref{non_boost}) in this language arises as a combination of a conventional linearly realized boost/rotation on the components of $X^\mu$, and a compensating gauge transformation restoring the gauge condition (\ref{Xmu}).
We deliberately chose a somewhat less elegant formulation, to stress the analogy with the more familiar case of Goldstones for internal symmetries, such as pions, and to make unitarity in the low energy effective theory manifest.

From a practical point of view  the NG term is often sufficient to describe the dynamics of a long string. However, higher order corrections may be of interest as well. In particular, they may become important for interpreting the continously improving lattice QCD data because the length of the flux tubes on the lattice are not that long compared to the  width of the string. It appears straightforward to incorporate these, using the action (\ref{coset_action}) and the standard effective field theory toolbox (see, e.g., \cite{Georgi:1994qn} for a nice and concise introduction). 

Before proceeding let us mention one possible subtlety which we are {\it not} going to address. The CCWZ proof of the uniqueness of the
non-linear realization does not appear to be directly applicable to space-time symmetries and, to the best of our knowledge, has not been extended to the case at hand. However, we find it rather implausible that physically inequivalent non-linear realization of the same coset $ISO(D-1,1)/ISO(1,1)\times SO(D-2)$ should exist.
 

A concern that a subtlety might be missed by the above effective field theory analysis comes from the theory of fundamental strings. In that case the goal is to quantize the Nambu--Goto action as a UV complete, rather than an effective theory. This turns out to be possible only in the critical number of dimensions, $D=26$ (in the absence of additional degrees of freedom on the worldsheet). Depending on the procedure, one encounters different pathologies away from $D=26$. In the light cone gauge one pays the price of losing the target space Lorentz symmetry, while  the ``old" covariant quantization introduces negative norm states in the physical Hilbert space. 

These results appear surprising from the effective field theory perspective. Nowhere in our previous discussion did we see a sign that something special happens at $D=26$, nor do we expect to have problems with Lorentz invariance and/or being able to construct a positive norm Hilbert space in simple quantum field theories such as an Abelian Higgs model giving rise to cosmic strings or in QCD at $D=4$. The main goal of our work from a theoretical viewpoint is to understand what is special about the critical dimension in the effective field theory description.

Of course, we are not the first to be puzzled by this. The issue has been addressed a number of times in the past, and has essentially been solved by Polchinski and Strominger (PS) in \cite{Polchinski:1991ax} (see also \cite{Polchinski:1992vg}). They chose to start with a fully covariant description of the string, and instead of working in static gauge make use of a conformal gauge in which the induced metric (\ref{hind}) is conformally flat.\footnote{Not to be confused with what one usually calls the conformal gauge in string theory. Here it is directly the induced metric which is fixed to be conformally flat, rather than the auxiliary Polyakov metric \cite{Polyakov:1981rd}, which is never introduced in the PS procedure.} This would probably not be the first choice for most effective field theorists. 
This gauge fixing leaves a huge residual gauge freedom. All conformal transformations  preserve the PS gauge condition. As a consequence, the theory in this gauge is not manifestly unitary. One is left with $D$ fields $X^\mu$ and has to impose constraints to restrict to the physical states. An advantage of this gauge is the linear realization of the Lorentz algebra. 

Instead of directly calculating the NG action in conformal gauge (including the contribution from the path integration measure) PS chose to let symmetries guide them. Conformal invariance fixes the form of the two leading terms in conformal gauge, leading to the following action,
\be
\label{PSaction}
S_{PS}=\int d^2\sigma\;\l-\frac{1}{2\ell_s^{2}}\l\d_\alpha X^\mu\r^2-2\beta {\l \d_{\alpha}\d_{\beta}X^\mu\d^\beta X_\mu\r^2 \over \left[\l\d_\gamma X^\nu\r^2\right]^2}+\dots\r\;.
\ee
One should not be scared by powers of $\l\d_\alpha X^\mu\r^2$ appearing in the denominator of this action. The long string background corresponds to taking, say
\[
X^0_\text{cl}=\tau\;,\;\;X^1_\text{cl}=\sigma\,.
\]
For the fluctuations around such a background, the PS action (\ref{PSaction}) becomes perfectly local. In other words, in the PS power counting only higher derivatives acting on $X^\mu$ are considered small, but not the first ones. This is related to the fact that the $X^\mu$'s are not identical with the physical degrees of freedom in this gauge.

A heuristic explanation of the PS term (the second one in (\ref{PSaction})) is that the Polyakov determinant for the auxiliary metric~\cite{Polyakov:1981rd}
\be
\label{Polyakov}
S_{Pol}={26-D\over 96\pi}\int d^2\sigma \sqrt{-h}R{1\over \Box}R
\ee
 takes exactly this form for the induced metric in conformal gauge. This identification suggests that the coefficient $\beta$ is fixed
 \be
 \label{beta}
 \beta={26-D\over 48\pi}\;.
 \ee
 Indeed, as argued by PS, the absence of ghosts in the spectrum of the conformal theory (\ref{PSaction}) fixes its central charge to be equal to 26, also resulting in 
 (\ref{beta}). This result has also been confirmed by an explicit calculation in a specific model in \cite{Natsuume:1992ky}.
 
 To summarize, the PS explanation of what is special about the critical dimension from an effective field theory viewpoint is very simple. In any number of dimensions, long strings are described by an effective action whose leading terms are given by (\ref{PSaction}), which is manifestly Lorentz invariant, and does not lead to ghosts. For $D=26$ the theory has a chance of being UV complete on its own. In any other number of dimensions, it is necessarily supplemented with a non-renormalizable PS term, suggesting that extra ingredients are needed for the UV completion. 

We feel that there are a number of interesting questions left to be understood. At the most basic level, they all reduce to understanding the PS argument in static gauge (\ref{Xmu}). In particular, when expanded around a long string background the PS term would lead to interactions among string perturbations
which do not naively follow from any of the local geometric invariants in (\ref{coset_action}). This is not surprising given that  the PS term corresponds to the non-local Polyakov determinant (\ref{Polyakov}). However, does this imply that the CCWZ construction is missing something and new non-geometric and/or non-local terms should be added in static gauge?

Apart from a natural theorist's desire to understand simple things in the most complicated ways there is a practical motivation to clarify these issues.
The quality of the lattice data reaches the point where subleading corrections to the NG action become important \cite{Teper:2009uf,Kuti:2005xg}, and historically most calculations of the corrections to the spectrum of the QCD string were performed in static gauge. 
%
The early calculation of the long string properties using the string effective action was done in \cite{Luscher:1980ac}. By using the free part of the NG action
\be
\label{Sfree}
S_{free}=-{1\over 2\ell_s^2}\int d^2\sigma\l\d_\alpha X^i\r^2\,,
\ee
the leading quantum correction to the energy of the string, the ``L\"uscher term", was calculated
\be
\label{Luscher}
\Delta E_L=-(D-2){\pi\over 6 R}\;,
\ee 
where $R$ is the length of the string. In conventional field theory language this correction is the one-loop Casimir energy arising as a result of compactification of the free two-dimensional theory (\ref{Sfree}). This correction is universal for all string states. The first corrections which distinguish different excited string levels were calculated much later in \cite{Luscher:2004ib}. They arise from the NG quartic interactions, which are of the form
 \be
 \label{quartic_NG}
 S_2+S_3=-{1\over 4\ell_s^2}\int d^2\sigma\l c_2\l\d_\alpha X^i\d^\alpha X^i\r^2+c_3\d_\alpha X^i\d_\beta X^i\d^\alpha X^j\d^\beta X^j\r\,.
 \ee 
In the NG action the coefficients are 
\be
\label{c23}
c_2=\frac12\;,\;\;c_3=-1\,,
\ee
and, as emphasized recently in \cite{Aharony:2009gg,Aharony:2010db,Aharony:2010cx,Aharony:2011gb}, this is the only choice compatible with the Lorentz symmetry of QCD.
The same group of authors also initiated the next order calculation of the corrections to the effective string spectrum.
It was suggested that to accomplish this the following term has to be introduced into the effective string action in static gauge
\be
\label{c4}
S_4=-c_4\int d^2\sigma \d_\alpha\d_\beta X^i\d^\alpha\d^\beta X^i\d_\gamma X^j \d^\gamma X^j\,,
\ee
and motivated by the PS result it was conjectured that $c_4=(D-26)/192\pi$.
Related to our earlier confusion how to reproduce the PS result in static gauge, this proposal appears surprising, because, as straightforward to check, the interaction (\ref{c4}) is not invariant under the non-linearly realized boosts (\ref{non_boost}). So another goal of our paper will be to (dis)prove this conjecture, and to set up  the correct procedure for calculating the spectrum of the effective strings at this order in static gauge.

Let us finish this review of the history of the subject and our goals with an outline for the rest of the paper and a brief summary of our own results.
In section~\ref{tree_level}, we explain  how to look for the PS term in static gauge. To this end, we will study the structure of the on-shell $2\to 2$ scattering amplitude of the string excitations. From the PS result and from the analysis in light cone gauge we know that the theory in $D=26$ does not exhibit annihilation, {\it i.e.} string excitations corresponding to oscillations in one direction cannot produce oscillations in one of the orthogonal directions. So the annihilation, corresponding to the PS term, should be proportional to $(D-26)$.
We check this logic at tree level, and find that it singles out the NG choice (\ref{c23}) even if one drops the assumption of the non-linearly realized Lorentz symmetry. 

In section~\ref{1loop_infinite}, we proceed to one-loop and study the divergent part of the $2\to 2$ scattering amplitude in NG. To preserve the non-linearly realized Lorentz symmetry we use dimensional regularization and find that the action (\ref{coset_action}) has to be supplemented  with the following counterterm,
\be
\label{Einsteinct}
S_{E}=-{D-8\over 48\pi\epsilon}\int d^{2-2\epsilon}\sigma\; \sqrt{-h} R\;.
\ee
As far as we can tell this is the only peculiarity in implementing the CCWZ program for the effective string at the quantum level. The theory contains a so-called evanescent operator, the Einstein-Hilbert term, which is a total derivative in $D=2$, but is required to be included in the action in $D=2-2\epsilon$ for the consistent renormalization of the theory. It vanishes at tree-level but may contribute to physical observables through loops. This subtlety does not compromise the CCWZ logic. On the contrary, it may be considered as a self-consistency check. There are two possible tensor structures at this order (the second one corresponds to the $c_4$-term (\ref{c4})). Only the one consistent with the non-linearly realized Lorentz symmetry appears in the counterterm.

In section~\ref{1loop_finite}, we inspect the finite part of the one-loop amplitude for $2\to 2$ scattering. As expected, the annihilation part of the amplitude 
 is proportional to $(D-26)$ and exactly agrees with the result derived from the PS action in conformal gauge. This confirms that the PS term has to be interpreted as the Polyakov determinant (\ref{Polyakov}). In static gauge it appears only as a part of the 1PI effective action, so its non-locality does not pose any problems.
It is special to conformal gauge that this term can be written in a local Lorentz invariant form and, as a result, appears directly in the Wilsonian action. To our knowledge this is the first derivation of the PS interaction by a direct calculation. Interestingly, this annihilation part of the amplitude is the most rapidly growing in the UV, suggesting that its cancellation may be required for renormalizability of the theory.

In section~\ref{Weyl}, we clarify the issue of the $c_4$-term (\ref{c4}). To make contact with  earlier works we work with a different regularization scheme from the previous sections, the one used in \cite{Aharony:2009gg,Aharony:2010db,Aharony:2010cx,Aharony:2011gb}.
The prescription is to use Weyl symmetric ordering with subsequent $\zeta$-function regularization.
The complication in this scheme compared to dimensional regularization is that it does not preserve the non-linearly realized Lorentz algebra.
We calculate the commutator of boosts in this scheme and show that the closure of the algebra requires the presence of the term~\eqref{c4} in the action with coefficient
\be
\label{truec4}
c_4=-{1\over 8\pi}\;.
\ee 
The appearance of non-covariant counterterms in certain regularization procedures is familiar from the early days of pion physics~\cite{Gerstein:1971fm} (see also~\cite{Cai:1994um} in the context of membranes in condensed matter systems), and has nothing to do with peculiarities of the string quantization {\it per se}. To reproduce the PS term in this gauge one should again calculate the $2\to 2$ scattering amplitude. This calculation is somewhat more laborious than the one we performed before, because in dimensional regularization a number of diagrams vanish, but contribute if one uses the Weyl ordering. In particular, the sixth order vertex from the NG action will contribute. We leave this calculation for future work.

In section~\ref{mass_spectrum}, we outline the calculation of the corrections to the string spectrum in static gauge at this order using the action derived here.

In the concluding section~\ref{conclusions}, we discuss future directions.

\section{Tree Level Warm-Up}
\label{tree_level}
As a warm up, let us explain our strategy for identifying the PS term in static gauge at tree level. 
It is clear from the PS derivation that this term is not an independent contribution to the action, but is an intrinsic part of the consistent quantization of the NG theory.
In conformal gauge it determines the leading string interactions not arising from the constraints. So we expect to be able to identify it by a careful inspection of the $2\to 2$ scattering amplitude. One may worry that these scattering amplitudes suffer from soft and collinear divergences and are not well-defined. In~\cite{critical} we will explain in detail that this worry is unjustified and the S-matrix exists even though this is a theory of massless particles. Both power counting and the origin of the PS term (it is supposed to arise from the gauge fixing determinant in conformal gauge) indicate that it should correspond to a one-loop effect in static gauge.


From the light cone quantization (which is consistent at $D=26$) we know the exact spectrum of the critical $D=26$ theory. After compactification on a circle\footnote{For definiteness and simplicity we consider periodic boundary conditions.} (see, e.g., \cite{Polchinski:1998rq}), 
\be
\label{LCspectrum}
E_{LC}(N,\tilde N)=\sqrt{{4\pi^2(N-\tilde{N})^2\over R^2}+{R^2\over \ell_s^4}+{4\pi\over \ell_s^2}\l N+\tilde{N}-{D-2\over 12}\r}\;.
\ee
Here $R$ is the length of the string, $N$ and $\tilde{N}$ are levels of an excited string state, so that $2\pi(N-\tilde{N})/R$ is the total Kaluza--Klein momentum of a state. 

To avoid a confusion, the following comment is in order. Often, the spectrum $E_{LC}$ is referred  to as the NG spectrum. This is correct in the critical number of dimensions, where the light cone quantization 
provides a solution of the NG theory and, in particular, preserves the target space Lorentz symmetry.
In any other number of dimensions, (\ref{LCspectrum}) is a spectrum of some different solvable Lorentz violating theory,\footnote{To be more precise, it is the target space Lorentz symmetry that is violated. The theory is still a relativistic field theory in $1+1$ dimensions.} and we find it a misnomer to call it the NG spectrum, because the NG theory is Lorentz invariant.
Instead, we will refer to (\ref{LCspectrum}) as the light cone spectrum.
 
Notice that this is not the spectrum of a free theory. The energies of $n$-particle states are not the sums of the energies of 1-particle states. We will discuss consequences of this simple observation in~\cite{critical}.
 A distinctive property of the spectrum (\ref{LCspectrum}) is that the energy is uniquely determined by the level of the state. This implies in particular that states at the same level, which belong to different representations of the ``flavor" group $SO(D-2)$ (the group of transverse rotations), are degenerate. Physically, this implies that a string oscillating in one direction does not start oscillating in other directions. So our guiding principle for identifying the critical number of dimensions and the PS interaction will be to look for  annihilations of, say, two $X^2$ quanta into two $X^3$ quanta. 

Let us first study the tree level $2\to 2$  scattering amplitude. In what follows, we will always consider on-shell amplitudes. However, keeping in mind the later use of dimensional regularization, we do not use its specific two-dimensional properties, unless stated otherwise.
In general, the $SO(D-2)$ flavor symmetry restricts the $2\to 2$ amplitude to take the form\footnote{We will use conventions in which $S_{\alpha\beta}=\mathbf{1}_{\alpha\beta}+i(2\pi)^d\delta^d(p_\alpha-p_\beta)\mathcal{M}_{\alpha\beta}$. In these conventions, the optical theorem implies that the imaginary part of the Feynman amplitude for forward scattering is positive definite.}
\be
\label{amplitude}
{\cal M}_{ij,kl}=A\delta_{ij}\delta_{kl}+B \delta_{ik}\delta_{jl}+C \delta_{il}\delta_{jk}\;,
\ee
%
%
where $i,j,k,l$ are the flavor labels of scattering particles.
Crossing symmetry implies that the amplitudes $A$, $B$ and $C$ satisfy the following relations as 
 functions of the Mandelstam variables $s$, $t$, and $u$
\begin{gather}
A(s,t,u)=A(s,u,t)=B(t,s,u)=C(u,t,s)\,.
\label{crossings}
\end{gather}
Our convention in defining the $s,t,u$ variables is that for an $i$-particle carrying momentum $p_1$, a $j$-particle carrying $p_2$, a $k$-particle carrying $p_3$ and an $l$-particle carrying $p_4$ 
\be
s=-(p_1+p_2)^2\,,\qquad t=-(p_1-p_3)^2\,\qquad\text{and}\qquad u=-(p_1-p_4)^2\,.
\ee
The absence of annihilations is simply the statement
that for the critical string the whole amplitude (\ref{amplitude}) is proportional to unity
\[
{\cal M}_{ij,kl}\propto \delta_{ik}\delta_{jl}\delta(p_1^\sigma-p_3^\sigma)\delta(p_2^\sigma-p_4^\sigma)+\delta_{il}\delta_{jk}\delta(p_1^\sigma-p_4^\sigma)\delta(p_2^\sigma-p_3^\sigma)\,.
\]
In particular, this condition implies $A=0$. Naively, the crossing relations (\ref{crossings}) then imply that there is no non-trivial scattering at all. As we will see momentarily, this argument fails in two space-time dimensions. It is a peculiarity of the two dimensional kinematics that either  $t=0$ and $u=-s$ (``$t$-channel"), or $u=0$ and $t=-s$ (``$u$-channel"). In other words, in two dimensions the absence of annihilations
\be
\label{A=0}
A_{d=2}=0
\ee
allows for a non-trivial $S$-matrix if $A\propto ut$. Moreover, as soon as the condition (\ref{A=0}) is satisfied and the amplitude in a general number of space-time dimensions $d=2-2\epsilon$
satisfies the crossing relations (\ref{crossings}), ${\cal M}_{ij,kl}$ is automatically proportional to unity at $d=2$. Note that pieces in $A$ proportional to $ut$ still carry physical information, because by crossing they allow to reconstruct parts
of $B$ and $C$ amplitudes, which are non-vanishing at $d=2$.

To see how all this works let us study  the tree level scattering  amplitudes  for a general choice of $c_2$, $c_3$ coefficients in the action (\ref{quartic_NG}).
\begin{figure}[t!] 
 \begin{center}
 \includegraphics[width=0.55in]{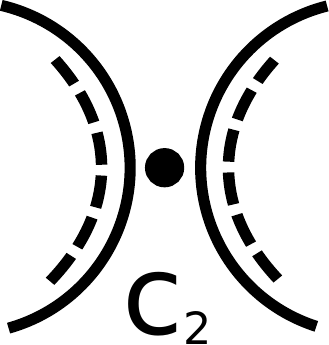}
 \hspace{40pt}
 \includegraphics[width=0.55in]{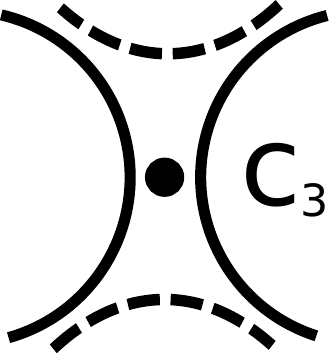}
 \caption{Quartic treel-level vertices following from the NG action. Solid lines follow the flow of flavor indices, and dashed lines show the contractions of momenta at the vertex.}
 \label{vertices}
 \end{center}
\end{figure}
With the one-loop calculation in mind, it is convenient to represent different vertices originating from~(\ref{quartic_NG}) as shown in Fig.~\ref{vertices}, where solid lines follow the flow of the flavor indices, and dashed lines show the momentum contractions. Using $s+t+u=0$, the amplitude for annihilation can be written as
\be
A=-{\ell_s^2\over 4}((c_3+2c_2)s^2-2c_3tu)\,,
\ee
and $B$ and $C$ can be obtained by the crossing relations (\ref{crossings}).
We find  that the relation (\ref{A=0}) corresponds to the NG choice
\[
2c_2+c_3=0\;.
\]
Note that the absolute values of $c_2$, $c_3$ can be rescaled by redefinition of the fields $X^i$ and $\ell_s$. The overall sign is fixed by the positivity (subluminality) constraint on the forward scattering amplitude \cite{Adams:2006sv}
\be
\label{forwardM}
{\cal M}_{forward}=\ell_s^{2}c_2s^2\,.
\ee
So at quartic level, the NG action in two dimensions is uniquely determined if one requires the shift symmetry and the absence of annihilations, without explicit reference to nonlinearly realized boosts and takes the form
\be
\label{Mtree}
{\cal M}_{ij,kl}=-\frac12\ell_s^2(\delta^{ik}\delta^{jl}su+\delta^{il}\delta^{jk}st)\,.
\ee

Let us now turn to the PS interaction. In general the physical states in the PS gauge are not merely the excitations of the $X^i$ components. However, for $2\to 2$ scattering at leading order in the PS interaction this is still the case \cite{Aharony:2011ga}. The leading interaction at low energies then arises through the constraints (see {\it e.g.} $(13)$ in~\cite{Polchinski:1991ax}) and agrees with~\eqref{Mtree} as expected. At the next order the PS term leads to a flavor changing contribution to $2\to 2 $ scattering
\be
\label{PS_amplitude}
{\cal M}^{PS}_{ij,kl}=-{D-26\over 192 \pi}\ell_s^4\l \delta^{ij}\delta^{kl}s^3+\delta^{ik}\delta^{jl}t^3+\delta^{il}\delta^{jk}u^3\r\;.
\ee
We see that away from the critical dimension a string initially oscillating in one plane may start to oscillate in a different one.
 In the next section we will reproduce this result directly in static gauge.  
\section{One-loop $2\to 2$ Scattering}
Studying the $2\to 2$ scattering amplitude to order $s^3$ will be interesting for two reasons. The infinite part of the amplitude will provide an explicit consistency check that the renormalization of the theory is compatible with the non-linearly realized Lorentz symmetry implemented through the CCWZ procedure, and the finite part of the one-loop amplitude in static gauge will reproduce the PS result~(\ref{PS_amplitude}).

To deal with the UV divergences 
we need to chose a regularization scheme. To preserve the manifest invariance under the non-linearly realized Lorentz symmetry, our choice should respect this symmetry. Fortunately, the favorite effective field theorist's choice, dimensional regularization, works yet again. The CCWZ action in any number $d=2-2\epsilon$ of (world-sheet) dimensions is invariant under the non-linearly realized $ISO(D-1-2\epsilon,1)$. Symmetry transformation rules do not depend on $\epsilon$, so that the invariance should hold order by order in $\epsilon$-expansion, and  the minimal subtraction scheme is compatible with the symmetry. This argument essentially proves that the CCWZ construction holds at the quantum level, but let us see how it works explicitly at one-loop.
\subsection{Infinite Part of the $2\to 2$ Amplitude and the Evanescent Einstein Term}
\label{1loop_infinite}

\begin{figure}[t!] 
 \begin{center}
 \includegraphics[width=2.8in]{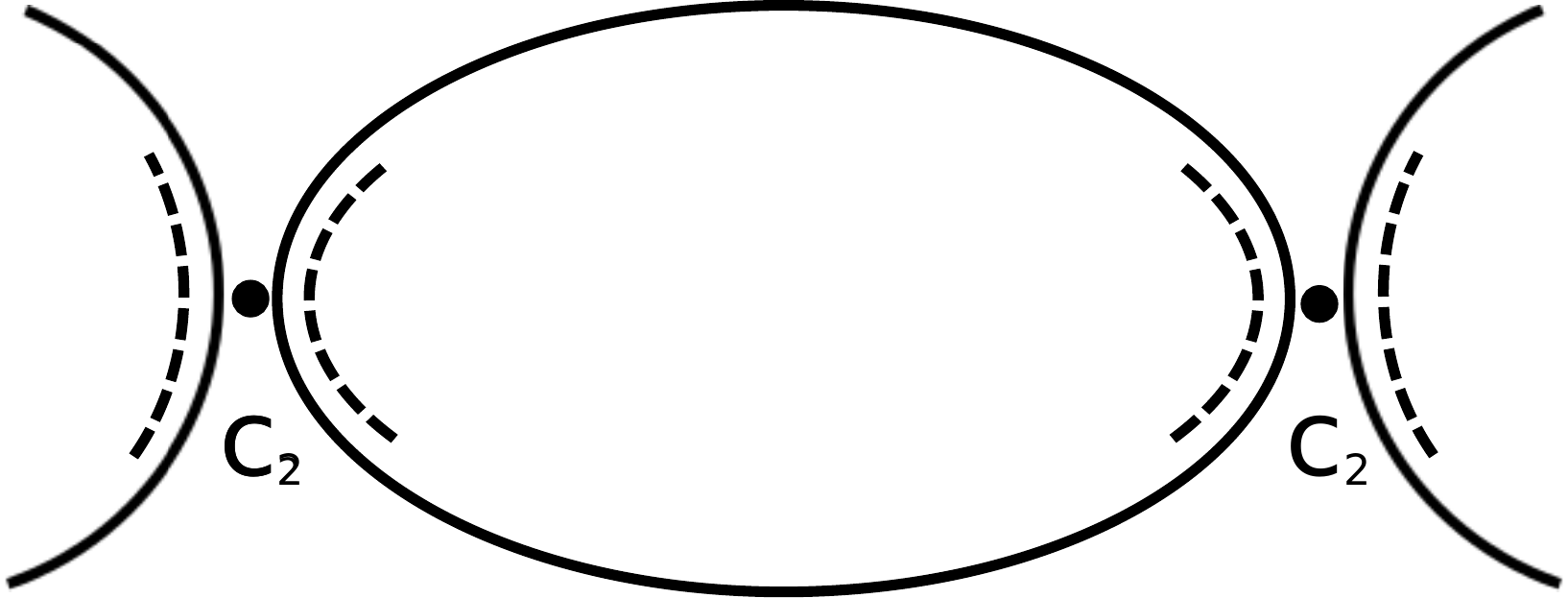}
 \hspace{40pt}
 \includegraphics[width=2.8in]{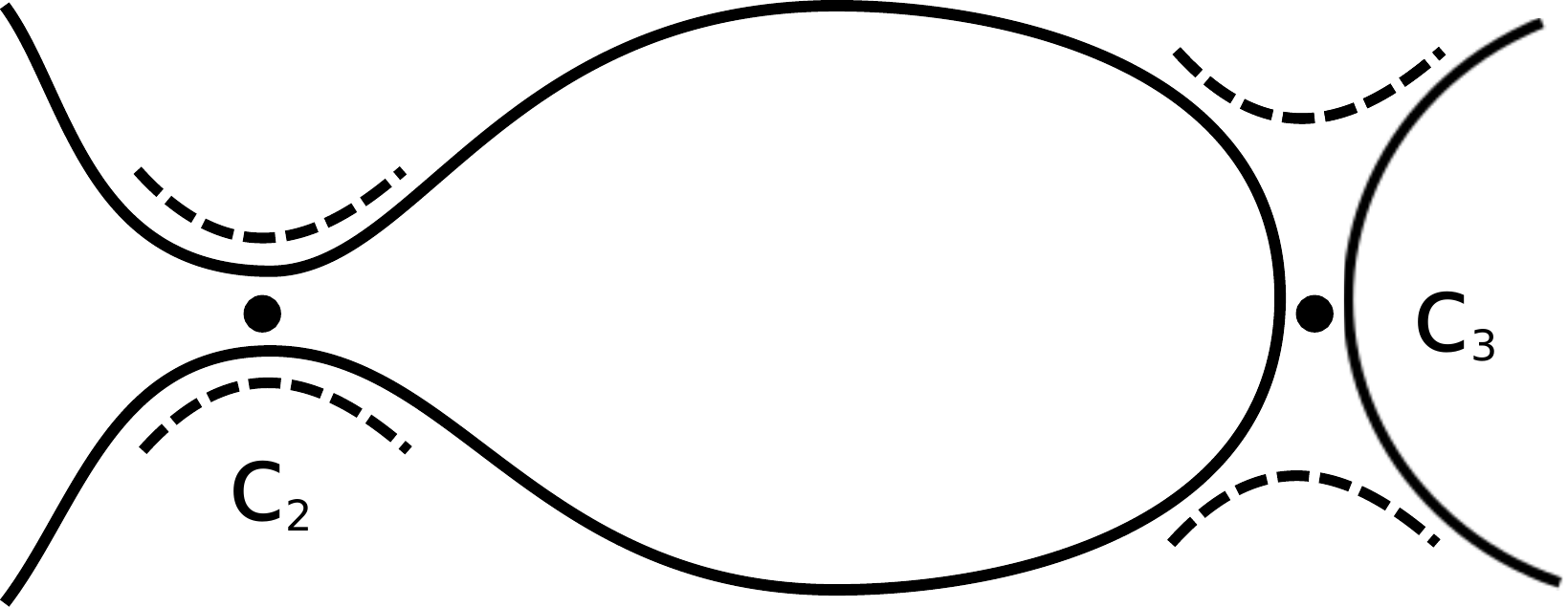}
   \put(-360,-15){a)}
  \put(-100,-15){b)}
 \caption{Sample one-loop diagrams contributing to the $2\to 2$ scattering in the NG theory. Solid and dashed lines have the same meaning as in Fig.~ \ref{vertices}.}
 \label{fig:loop}
 \end{center}
\end{figure}

The rigidity term in (\ref{coset_action}) does not contribute to the on-shell amplitude at the $s^3$ level, and the scattering at this order is determined by the NG action. 
In dimensional regularization it is given by the ``fish" diagrams coming from the $c_2$ and $c_3$ vertices.
Accounting for all possible contractions of the flavor and space-time indices results in a surprisingly large number of Feynman diagrams. Figure~\ref{fig:loop} shows some representative examples. As seen from these examples,
there are two classes of diagrams, those 
 with 
a closed loop of the flavor flow  like the one on  Fig.~\ref{fig:loop}a and those without such a loop, as the one in  Fig.~\ref{fig:loop}b. 
The contribution of the diagrams of the first kind is proportional to $(D-2)$, while the second class
of diagrams produces the $D$-independent result.
 The presence of these two different topologies opens a room for a special value of $D$.

We encounter a little puzzle here that was already mentioned in the {\it Introduction}.. There is no tree-level vertex that is compatible with the non-linearly realized Lorentz symmetry and contributes to the on-shell $2\to 2$ scattering at order $s^3$. However, the diagrams presented in Figure~\ref{fig:loop} lead to logarithmic divergences and require a counter-term.
The puzzle is resolved by the presence of an additional operator, the Einstein-Hilbert term $\sqrt{-h}R$, omitted in our initial CCWZ action (\ref{coset_action}). The reason for the omission is that 
it becomes a total derivative at the physical number of space-time dimensions $d=2$. As a result it does not contribute to the tree-level scattering and a change of the coefficient in front of this operator by a finite amount can be compensated by a change of the subtraction scheme \cite{Dugan:1990df}. Nevertheless, this operator has to be included in dimensional regularization for a consistent renormalization
of the theory, and the infinite part of the corresponding coefficient is fixed unambigiously. When inserted in loops, this operator contributes to the physical observables. 

 To see that adding this operator is enough to remove all the divergences at this order, note
that in general at the level of four fields and six derivatives there are two linearly independent 
$ISO(1,1)\times SO(D-2) $ invariant local interaction vertices which do not vanish on-shell in $d$ world-sheet dimensions,
\be
\d^{\beta}X^j\d^{\gamma}X^i  \d_{\alpha}\d_{\beta}X^i\d^{\alpha}\d_{\gamma}X^j      \qquad \text{and}\qquad \l \d_{\alpha}\d_{\beta}X^i\d^\beta X^i\r^2 \;.
\ee
The first term appears in the expansion of the Einstein-Hilbert term, which up to total derivatives is
\be
\label{Einstein_term}
\int d^d \sigma\;\sqrt{-h}R=\int d^d\sigma\;  \d^{\beta}X^j\d^{\gamma}X^i\l  \d_{\alpha}\d_{\beta}X^i\d^{\alpha}\d_{\gamma}X^j-\d_\beta\d_\gamma X^i \Box X^j\r+\dots
\ee 
(the second term in (\ref{Einstein_term}) vanishes on-shell).
This interaction leads to a contribution to the scattering amplitudes of the form (\ref{amplitude}) with 
\be
\label{EinA}
A=-\frac{1}{2}\ell_s^4stu\,.
\ee 
The second term is the non-covariant $c_4$-term~\eqref{c4} and leads to an amplitude of the form~(\ref{PS_amplitude}).
 Non-linearly realized Lorentz-invariance predicts that all the divergences
in the $2\to 2$ scattering should be of the form (\ref{EinA}), so that the divergences can be canceled  at this order by inclusion of the evanescent Einstein-Hilbert term.
Calculation of the infinite part of the amplitude gives
\be
A= 
  \frac{\ell_s^4}{32\pi} \l{1\over \eps}-\gamma_E+\log 4\pi \r\left ({D\over 2}\l 2c_2+c_3\r^2 s^3- 
                                \frac{1}{3}\l Dc_3^2-2c_2^2-22c_2 c_3-\frac{37}{2}c_3^2\r stu \right)\;.
\ee
In agreement with the above expectation the divergence proportional to $s^3$ cancels for the Lorentz-invariant choice of the $c_2$ and $c_3$ coefficients.
This provides an explicit one-loop consistency check of the Lorentz-preserving properties of our regularization.
For the NG values (\ref{c23}) of $c_2$ and $c_3$ we find 
\be
A=-{D-8\over 96\pi}\ell_s^4 stu \l{1\over \eps}-\gamma_E+\log 4\pi \r
\ee
which gets cancelled by a Lorentz-covariant counterterm of the form (\ref{Einsteinct}).
\subsection{Finite Part of the $2\to 2$ Amplitude and Polchinski--Strominger Action}
\label{1loop_finite}
Let us now inspect the finite part of the amplitude.
A slightly lengthy but straightforward calculation provides the following finite one-loop contribution to the coefficient $A$ for the NG choice of the $c_2$ and $c_3$ coefficients
\begin{gather}
\label{finiteA}
 A=  -\frac{\ell_s^4}{192 \pi}\l  (D-26)s^3 + 
  {stu}\l \frac{16}{3}D+\frac{4}{3}-2(D-8)\log{-s\over \mu^2} \r  
   +12 tu\l t \; \log{s\over t}+u \; \log{s\over u}\r \r
\end{gather}
The amplitude consists of three pieces. The first $s^3$ term in (\ref{finiteA}) is the only one which does not vanish at $d=2$. It exactly matches the annihilation
amplitude (\ref{PS_amplitude})  from the PS action in conformal gauge. Consequently, this calculation provides an explicit derivation of the PS action from the static gauge point of view. As it should be,  conformal and static gauge provide the same result for the physical amplitudes. The only subtlety is that in conformal gauge the  PS interaction already appears at the level of the Wilsonian action, but cannot be presented in the local covariant form 
 in static gauge. As a result, in static gauge it arises in the 1PI action, which is not required to be local.
 
 The second term in (\ref{finiteA}) has the same structure as the contribution coming from the evanescent Einstein-Hilbert term (\ref{EinA}). It vanishes identically at $d=2$, by which we mean that the corresponding terms related by crossing in $B$ and $C$ amplitudes (see (\ref{amplitude})) are also zero. This is the only part of the amplitude which depends on the renormalization scale $\mu$, so that the physical on-shell amplitude is RG invariant as it should be.
 
 Finally, the remaining logarithmic terms in (\ref{finiteA}) are proportional to $ut$ and do not contribute on-shell in the annihilation channel at $d=2$.
By crossing they give rise to non-vanishing $B$ and $C$ amplitudes similarly to what happens at tree level. 
For instance, the on-shell $d=2$ $B$-amplitude is equal to\footnote{We restored the relevant $i0^+$ from the propagators.}
\begin{gather}
\label{finiteB}
 B_{d=2}= -\frac{\ell_s^4}{192 \pi}\l  (D-26)t^3 +12 su\l s \; \log{t\over s+i 0^+}+u \; \log{t\over u+i0^+}\r \r
\end{gather}

As usual the presence of the logarithmic terms is required by unitarity. A peculiar property of two-dimensional kinematics is that even though these logarithms do provide the necessary imaginary parts in the $B$ and $C$ amplitudes, the real logarithmic part vanishes on-shell at $d=2$, and the whole amplitude is purely polynomial. Indeed,  the $\log$ terms in (\ref{finiteB}) vanish in the $u$-channel ($u=0$ and $t=-s$) and in the $t$-channel ($t=0$ and $u=-s$) reduce to 
\be
\label{oneloopB}
B=i\frac{\ell_s^4}{16}s^3\,.
\ee
We see explicitly that no IR divergences arise at this order.

It is worth remarking that in $D=3$ the PS interaction gives rise to an amplitude proportional to $s^3+t^3+u^3$. This vanishes on-shell in $d=2$ and the interaction is absent in this case as it should be.
\section{Lorentz  Algebra of Weyl Ordered Effective Strings}
\label{Weyl}
So far we have worked in $2-2\epsilon$ dimensions to regulate the theory because this manifestly preserved the Poincar\'e symmetry of the $D$-dimensional space-time. We now discuss a different (non-covariant) procedure that is commonly used in the quantization of strings and was recently used for effective strings in~\cite{Aharony:2009gg,Aharony:2010db,Aharony:2010cx,Aharony:2011gb}. We compactify the theory by indentifying $\sigma\sim\sigma+R$. Consequently, we also compactify one of the space-time directions, $X^1\sim X^1+R$, and describe a closed string wrapping this compact direction once. We define all composite operators in terms of Weyl-ordered products and use $\zeta$-function regularization to ensure the finiteness of their matrix elements on physical states. 

From the quantization of the fundamental string in light-cone gauge, we know that this prescription generically only preserves the manifest $SO(D-2)$ symmetry, with the full Lorentz algebra closing only in $D=26$ dimensions. 

In static gauge the situation is somewhat similar. In this case the commutator of broken boosts $J^{0i}$ and $J^{0j}$ derived from the covariant action does not close on the unbroken $SO(D-2)$-generator $J_{ij}$ in the quantum theory, but now {\it independent} of the number of dimensions. It fails to do so, of course, because of terms that are suppressed by powers of $\hbar$ (which we will restore in this section) and $\ell_s$. In order for the quantum theory to be Lorentz invariant, one must thus add non-covariant terms to the action. The original charges then no longer leave the action including the non-covariant terms invariant. However, the charges can be modified so that they do generate symmetries of the modified action. As a consequence the classical algebra no longer closes, but the operators of the quantum theory do satisfy the Lorentz algebra. 

We will now show this explicitly at one loop for the term ~\eqref{c4}, which is the leading term in the $\ell_s$ expansion of an infinite number of non-covariant terms that have to be added in this regularization scheme. We begin by calculating the commutator of the boosts derived from the covariant action to order $(\ell_s/R)^2$. 
The broken boost generators take the form
\be
J^{0i}=\int\limits_{0}^{R}d\sigma\,\tau\Pi^i-X^i\mathcal{H}\,.
\ee
The second term is taken to be Weyl ordered and $\mathcal{H}$ is the Hamiltonian density of the theory. Working in an expansion in $1/R$, it is of the form
\be
\mathcal{H}=\frac{1}{\ell_s^2}+\mathcal{H}_2+\mathcal{H}_4+\cdots\,.
\ee
The quadratic Hamiltonian
\be
\mathcal{H}_2=\frac12\left(\ell_s^2{\Pi^i}{\Pi^i}+\frac{1}{\ell_s^2}{X^i}'{X^i}'\right)\,,
\ee
is of order $1/R^2$, $\mathcal{H}_4$ of order $\ell_s^2/R^4$, and the omission stands for terms of order $\ell_s^4/R^6$ and higher. In the interaction picture with $\mathcal{H}_\text{free}=1/\ell_s^2+\mathcal{H}_2$, the fields enjoy the mode expansion
\bea
X^i&=&x^i+\ell_s^2\frac{p^i}{R}\tau+\frac{i\ell_s}{\sqrt{4\pi}}\sum\limits_{n\neq 0}\left[\frac{1}{n}\alpha^i_n e^{-\frac{2\pi i n}{R}(\sigma+\tau)}+\frac{1}{n}\tilde\alpha^i_n e^{\frac{2\pi i n}{R}(\sigma-\tau)}\right]\,,\\
\Pi^i&=&\frac{p^i}{R}+\frac{\sqrt{\pi}}{\ell_s R}\sum\limits_{n\neq 0}\left[\alpha^i_n e^{-\frac{2\pi i n}{R}(\sigma+\tau)}+\tilde\alpha^i_n e^{\frac{2\pi i n}{R}(\sigma-\tau)}\right]\,,
\eea
where the non-vanishing commutation relations among the operators $x^i$, $p^i$, $\alpha^i_n$ and $\tilde\alpha^i_n$ are
\be
\left[x^i,p^j\right]=i\hbar\delta^{ij}\,,\qquad\left[\alpha^i_n,\alpha^j_m\right]=n\hbar \delta^{ij}\delta_{n+m,0}\qquad\text{and}\qquad\left[\tilde\alpha^i_n,\tilde\alpha^j_m\right]=n\hbar\delta^{ij} \delta_{n+m,0}\,,
\ee
and the operator $\mathcal{H}_4$ is 
\be
\mathcal{H}_4=\frac12\ell_s^2({\Pi^i}{X^i}')^2-\frac{1}{8}\ell_s^6({\Pi^i}{\Pi^i})^2-\frac{1}{8}\frac{1}{\ell_s^2}({X^i}'{X^i}')^2-\frac{1}{4}\ell_s^2{\Pi^i}{\Pi^i}\,\,{X^i}'{X^i}'\,.
\ee
Notice that the rigidity term does not contribute at this (or the next) order because it is proportional to the equations of motion for the free fields. With this Hamiltonian density and the mode expansion, a cumbersome but straightforward calculation yields the commutator of two boosts 
\be\label{eq:comm}
\left[J^{0i},J^{0j}\right]=-i\hbar\int\limits_{0}^{R}d\sigma\,X^i\Pi^j-X^j\Pi^i+\frac{i\hbar^2\ell_s^2}{2\pi}\int\limits_{0}^{R}d\sigma\,{X''}^i\Pi^j-{X''}^j\Pi^i\,.
\ee
The first term is the desired unbroken $SO(D-2)$-generator
\be
J^{ij}=\int\limits_{0}^{R}d\sigma\,X^i\Pi^j-X^j\Pi^i\,,
\ee
and the second term shows that in this regularization scheme the quantum theory for the covariant action~\eqref{coset_action} is not Lorentz-invariant. So non-covariant terms, at this order the term~\eqref{c4}, must be added with appropriate coefficients to obtain a Lorentz invariant quantum theory. To see how this works explicitly, note that the variation of (\ref{c4}) under the transformation (\ref{non_boost}) is
\bea
\updelta^{\alpha j}_\eps S_4= \int d^2 \sigma\,\partial^2X^kF_\eps^{kj\alpha}[X]+\partial_\gamma G_\eps^{\gamma\alpha j}[X]+\dots\,,
\eea
where the dots stand for terms higher order in fields and derivatives,
\be
F_\eps^{kj\alpha}[X]=2\eps c_4\left(\partial^2X^k \partial^\alpha X^j+\partial_\beta X^k\partial^\alpha\partial^\beta X^j-\delta^{jk}\partial_\beta X^i\partial^\alpha\partial^\beta X^i\right)\,,
\ee
and the explicit form of $G_\eps^{\gamma\alpha j}[X]$ in the total-derivative term is not needed as long as we are only interested in the modification of the transformation rule of the fields $X^i$.
We see that the non-covariant term~\eqref{c4} varies into a total derivative and a term proportional to the lowest order equations of motion.\footnote{The decomposition into the first and second term is not unique, but the ambiguity does not affect the transformation rules for the interaction picture fields.} 

This means we can define a new transformation $\bar\updelta^{\alpha j}_\eps$ under which the action including the term~\eqref{c4} varies into a total derivative. Under this transformation the quadratic part of the Nambu-Goto action varies into
\be 
 \bar\updelta^{\alpha j}_\eps S_\text{free} = {1\over \ell_s^2} \int d^2 \sigma\, \d^2 X^i \,\bar\updelta^{\alpha j}_\eps X^i-\partial_\gamma(\partial^\gamma X^i\bar\updelta^{\alpha j}_\eps X^i)\,.
\ee
So at this order, the new variation
\be
\bar\updelta^{\alpha j}_\eps X^k=\updelta^{\alpha j}_\eps X^k-\ell_s^2F_\eps^{kj\alpha}[X]\,, 
\ee
transforms the action including the Nambu-Goto terms and the term~\eqref{c4} into a total derivative. 
For the interaction picture fields, the first term in $F_\eps^{kj\alpha}[X]$ vanishes and the transformation for the boost takes the form
\be
\bar\updelta^{0j}_\eps X^k=-\eps\left(\tau\delta^{jk}-X^j\dot{X}^k-2c_4\ell_s^2\left[\partial_\beta X^k\partial^\beta \dot{X}^j-\delta^{jk}\partial_\beta X^i\partial^\beta \dot{X}^i\right]\right)\,.
\ee
The classical commutator for this transformation on interaction picture fields evaluates to
\be
\left[\bar\updelta^{0i}_{\eps_1},\bar\updelta^{0j}_{\eps_2}\right]X^k=-\eps_1\eps_2\left(X^j\delta^{ik}-X^i\delta^{jk}\right)+4\eps_1\eps_2c_4\ell_s^2\left({X''}^j\delta^{ik}-{X''}^i\delta^{jk}\right)+\dots\,,
\ee
where the omission stands for terms higher order in $\ell_s/R$. Independent of the number of space-time dimensions, at one loop the quantum theory will then be Lorentz-invariant provided
\be
c_4=-\frac{\hbar}{8\pi}\,,
\ee
because the modification of the classical algebra due to the term~\eqref{c4} precisely cancels the 1-loop contribution in~\eqref{eq:comm}. This is unrelated to the PS term. To reproduce it in this renormalization scheme, one should again calculate physical observables such as energy shifts or scattering amplitudes. The calculation is more involved than in dimensional regularization because more diagrams contribute. In particular, the sextic Hamiltonian is needed. 


\section{Energy Spectrum of Confining Strings}
\label{mass_spectrum}
As explained in the {\it Introduction} the major practical motivation for the study of effective strings comes from lattice QCD simulations, where the principal physical observables are the energies of the string excitations for finite length strings.
Let us briefly discuss the current status of these calculations and implications of our results. For concreteness and simplicity we will discuss closed strings with periodic boundary conditions. 
Currently, two different methods are being used for these calculations, using the static gauge theory and the PS theory, respectively.
We will mainly discuss the static gauge approach, which is the main focus of our paper.
 
From a phenomenologist's point of view this is a rather familiar problem of calculating the loop corrections to the spectrum of Kaluza--Klein modes (cf. \cite{Cheng:2002iz}, where this kind of calculation was performed for the Standard Model).
Compactification of a $(1+1)$-dimensional theory is however somewhat special. In particular, vacuum diagrams (Casimir energy), enter on equal footing with corrections to the excited KK state energies. 

\begin{figure}[t!] 
 \begin{center}
 \includegraphics[width=1.7in]{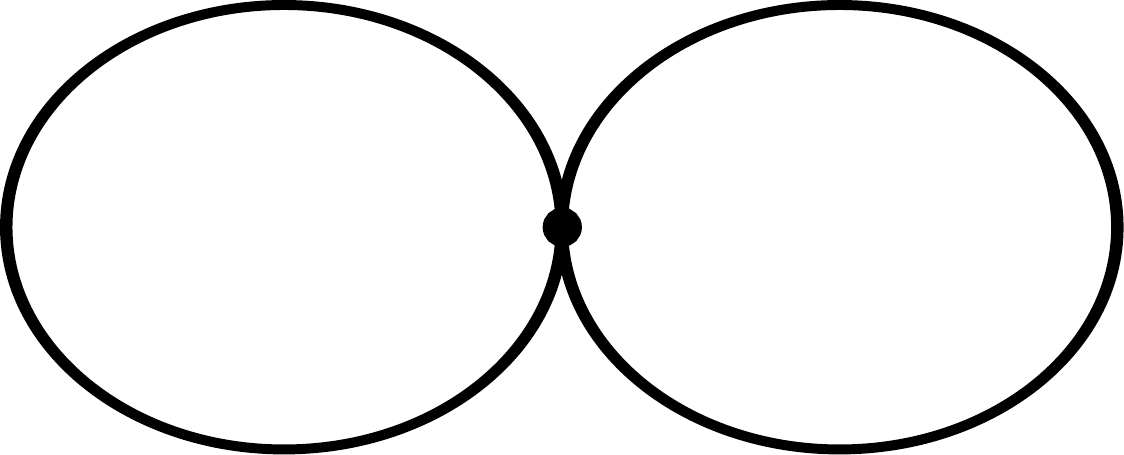}
 \hspace{20pt}
 \includegraphics[width=1.7in]{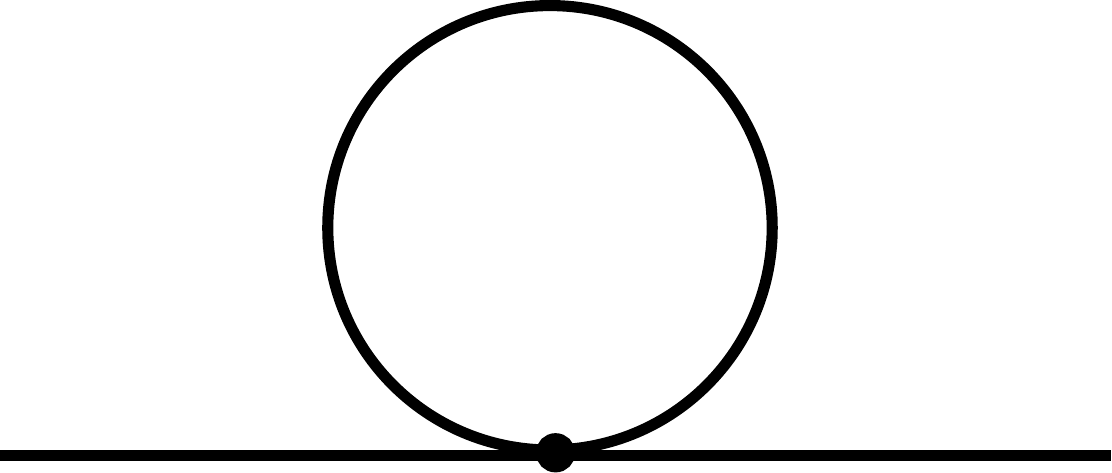}
  \hspace{20pt}
 \includegraphics[width=1.7in]{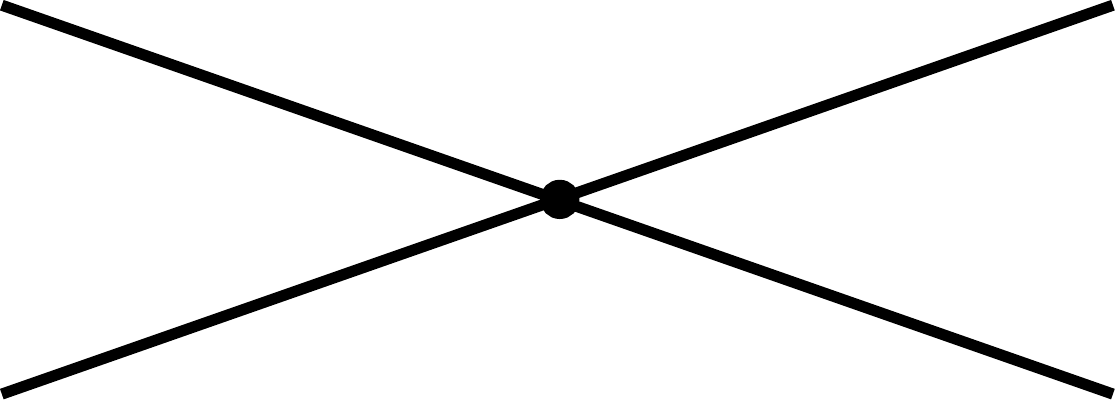}
 \caption{Connected diagrams contributing to the $R^{-3}$ order corrections to the energy levels. }
 \label{Rm3}
 \end{center}
\end{figure}

The classical ground state energy of the string is proportional to its length $R$. The classical energies of the excited states and the one-loop Casimir energy (the L\"uscher term  \cite{Luscher:1980ac}) scale as $R^{-1}$.
Order $R^{-3}$ corrections include two-loop vacuum graphs, one-loop corrections to one-particle excitations and tree-level corrections to two-particle excitations (see Fig.~\ref{Rm3}). At this order all the corrections come from
the Nambu--Goto part of the action.
These were calculated in \cite{Luscher:2004ib} using dimensional regularization. Note, that for these finite volume calculations one should include contributions from graphs which normally 
vanish in dimensional regularization, such as the vacuum and self-energy graphs in Fig.~\ref{Rm3}. This calculation (confirmed later in \cite{Aharony:2010db} using the Weyl symmetric ordering) demonstrated that at this order the spectrum of effective strings agrees with the light cone spectrum (\ref{LCspectrum}).

\begin{figure}[t!] 
 \begin{center}
 \includegraphics[width=1.7in]{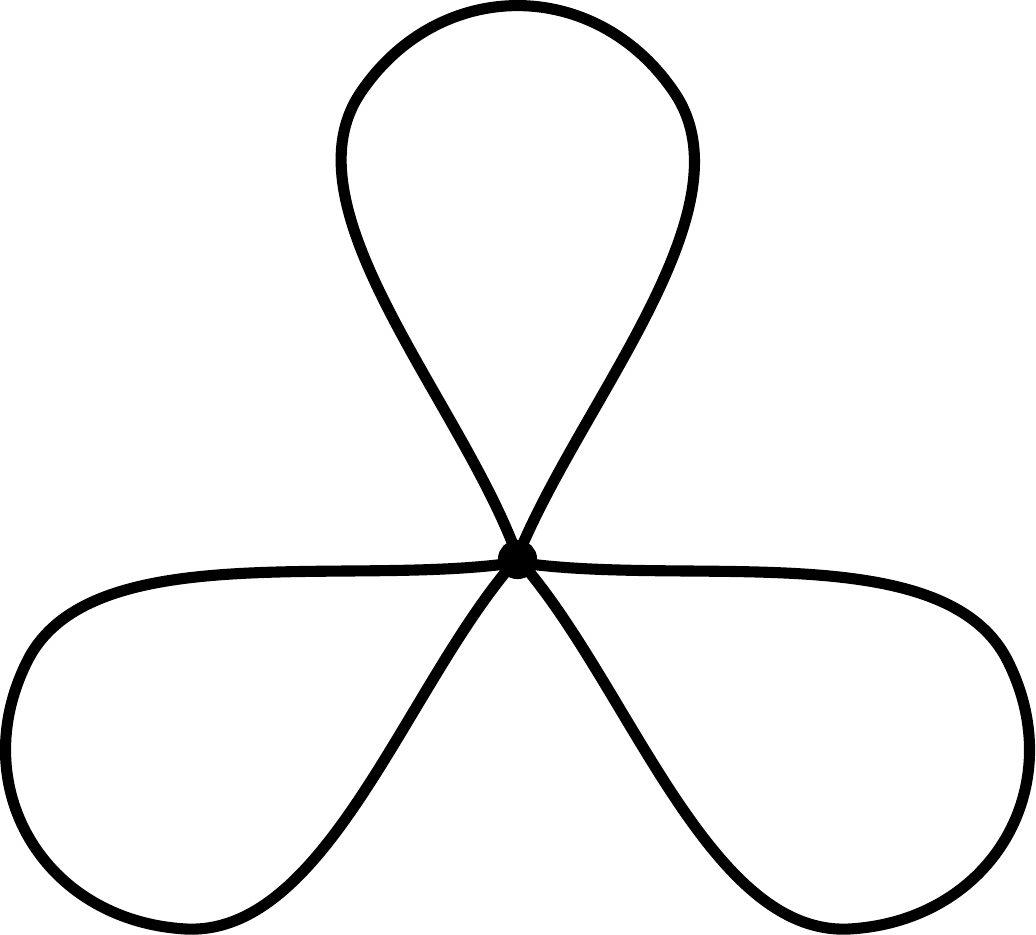}
 \hspace{20pt}
 \includegraphics[width=1.7in]{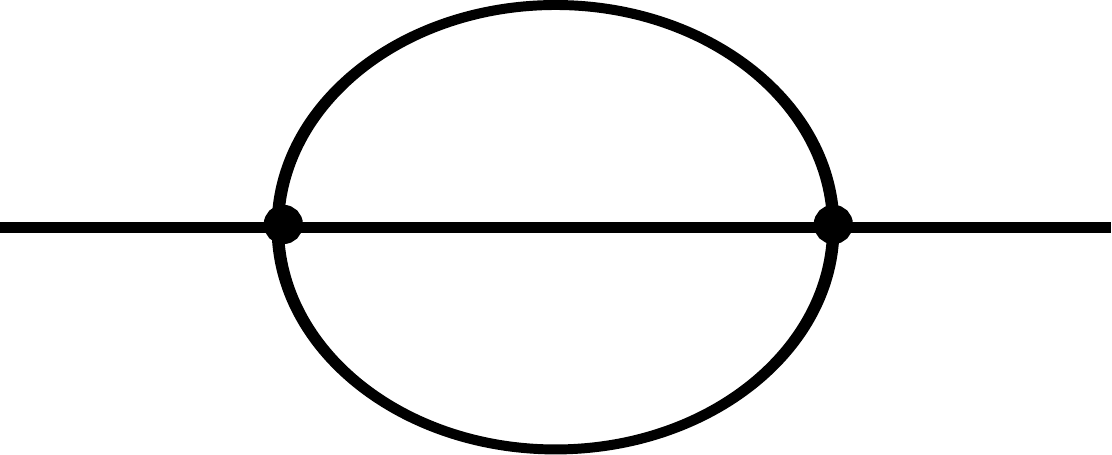}
  \hspace{20pt}
 \includegraphics[width=1.7in]{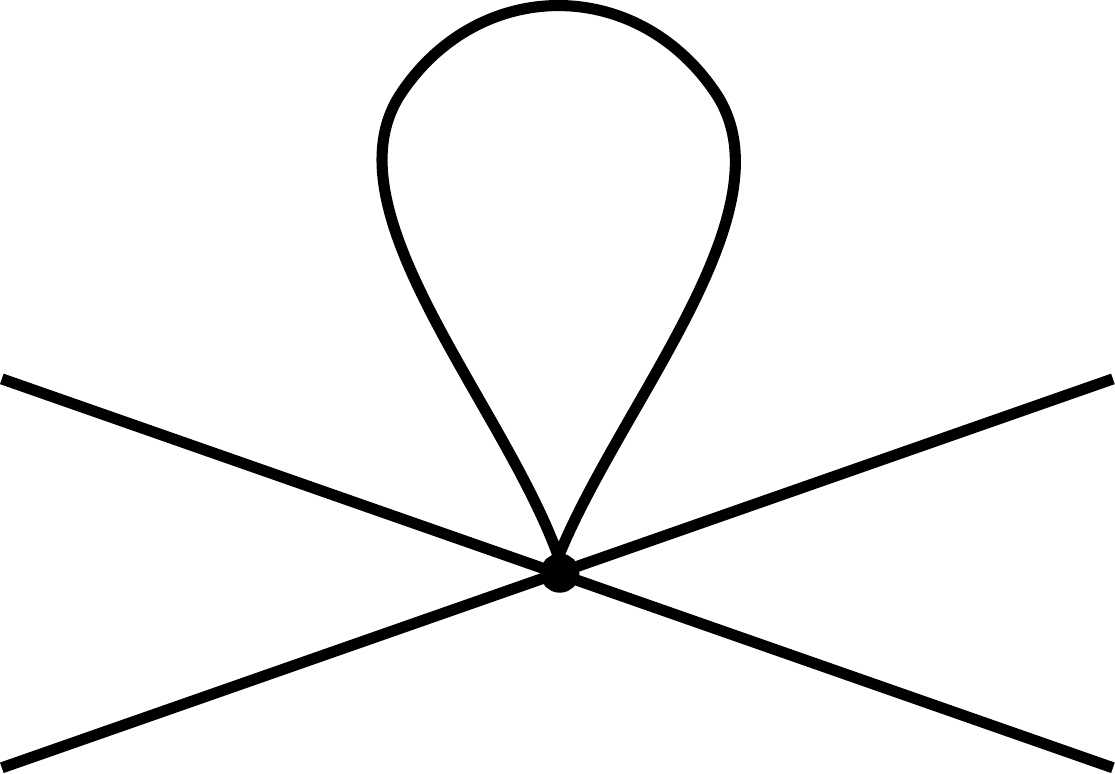}
 \caption{Sample connected diagrams contributing to the $R^{-5}$ order corrections to the energy levels.}
 \label{Rm5}
 \end{center}
\end{figure}

In principle, it is straightforward to proceed to higher orders in the $1/R$ expansion, using either dimensional regularization or Weyl ordering. At $1/R^5$ level all the corrections are still determined by the NG part of the action. A brute force calculation at this order  appears quite tedious, however. One needs to include the sextic interaction from the NG action.
Diagrams which have to be included at this order
include 3-loop Casimir contributions\footnote{These are universal for all levels, and can be ignored if one is only interested in the energy differences between the levels.},
2-loop corrections to self-energies, etc. (see Fig.~\ref{Rm5}). In dimensional regularization one also has to include  the evanescent Einstein-Hilbert term (\ref{Einsteinct}) which may contribute in loops at this order. With Weyl symmetric
ordering one has to include the contribution from the non-covariant counterterm (\ref{c4}) with the value of $c_4$ as given by (\ref{truec4}). 

An attempt to perform this calculation was made recently in \cite{Aharony:2010db}, using the Weyl symmetric ordering. It was assumed that all the diagrams without the $c_4$ term add up into the light cone spectrum (\ref{LCspectrum}) expanded up to this order. So rather than including all the diagrams above, only the contribution of the $c_4$ term was calculated. It turns out that the $c_4$ term contributes only to the tree level shift of two-particle states.
It was conjectured that the correct value of $c_4$ is $(D-26)/192\pi$, and this tree-level result was suggested as the leading correction to the light cone spectrum.

As we saw the correct value of $c_4$ is given by  (\ref{truec4}), and we see no reason for the diagrams with $c_4=0$ to reproduce the light cone spectrum\footnote{In fact, as will become clear momentarily they do not reproduce the light cone spectrum.},
so that this calculation is incomplete. 

However, here comes the puzzle. Recently the calculation at the $1/R^5$ order was performed in the PS gauge \cite{Aharony:2011ga} and yielded the same result as the one obtained in   \cite{Aharony:2010db}.
We should note in passing that the practical advantage of the PS gauge is that at this order one has to work with a free theory with a single interaction term. The price to pay is that one has to impose the BRST constraints to restrict to the correct physical states.
The puzzle is why the calculation in the PS gauge agrees with the incomplete one using the wrong value of $c_4$.

The explanation is as follows. Rather than doing the full brute force calculation in static gauge one can make use of the known light cone spectrum. This is the exact spectrum of some  
relativistic integrable two-dimensional theory. At this order in the derivative expansion its Lagrangian takes the form
\be
\label{light_cone}
{\cal L}_{LC}={\cal L}_{NG}+{D-26\over 192\pi}\d_\alpha\d_\beta X^i\d^\alpha\d^\beta X^i\d_\gamma X^j \d^\gamma X^j\;.
\ee
Here ${\cal L}_{NG}$ is the {\it full} renormalized NG action at this order in derivative expansion. For example, in dimensional regularization  ${\cal L}_{NG}$ includes the evanescent term (\ref{Einsteinct}), with Weyl symmetric ordering ${\cal L}_{NG}$ includes the non-covariant $c_4$-term with the correct value (\ref{truec4}) of $c_4$. The additional term
in (\ref{light_cone}) cancels the PS annihilation amplitude and breaks non-linearly realized Lorentz symmetry. 
The expression (\ref{light_cone}) immediately implies that the  leading $1/R^5$-difference between the Lorentz invariant NG spectrum and  the light cone spectrum is indeed determined by the matrix element of the $c_4$-operator with $c_4$ set to $(D-26)/192\pi$. 
This trick is of limited use for calculating the higher order corrections to the energy spectrum, and the $1/R^7$ calculation is likely to be quite laborious. 

To conclude this topic, let us mention one motivation to push these calculations one order further. At $1/R^7$ order the effective string spectrum becomes sensitive to the rigidity operator in (\ref{coset_action}). On the other hand, the rigidity coefficient $\alpha_0$ does not get renormalized in the effective theory. The leading
term in the expansion of the rigidity term is $(\Box X^i)^2$. It would be generated from the one-loop self-energy diagram, but this is zero in dimensional regularization.
Consequently, determination of the rigidity parameter of the confining strings would provide a direct probe of the underlying gauge dynamics.

\section{Future Directions}
\label{conclusions}
To summarize, in this paper we provided a derivation of the Polchinski--Strominger interaction in static gauge, which is arguably the
most  ``effective field theorist's" gauge to describe the dynamics of long strings. Admittedly, our derivation does not provide an
exhaustive answer to the question why the critical number of dimensions is special from the static gauge point of view.
 In particular, we did not discuss to what extent  the static gauge effective string theory becomes renormalizable at $D=26$ and did not say much about this theory.
 
 This deficiency will be fixed to some extent in the upcoming companion paper \cite{critical}, and this concluding section may be considered as a brief summary of our results there and an invitation to read \cite{critical}. The main idea, allowing to save a lot of work, is very simple and has already been exploited in the current paper for identifying the PS interaction. A great deal of information about critical string theory in the infinite volume limit can be extracted from its spectrum at finite volume, given by the light cone formula (\ref{LCspectrum}). The second observation is that in spite of being described by a massless two-dimensional theory, string excitations do not suffer from
 IR divergences and give rise to a well-defined S-matrix. 
 It follows from the light cone spectrum that this S-matrix is factorizable and  reflectionless and is entirely determined by the phase shift in the elastic $2\to 2$
 scattering.
 This phase shift can be determined from the known finite volume spectrum using standard lattice techniques \cite{Luscher:1986pf,Luscher:1990ck}. As a result one finds the following exact expression for the phase shift
 \be
\label{exactS}
\e^{2i\delta(s)}=\exp\l{i { \ell_s^2s/ 4}}\r\;,
\ee
in agreement with our perturbative results here and with the general form of the diagonal massless scattering, as determined by analyticity, unitarity and crossing-symmetry  \cite{Zamolodchikov:1991vx}.

Despite its simplicity, the answer (\ref{exactS}) is quite peculiar. It exhibits an essential singularity at $s=\infty$, even though the scattering amplitude (\ref{exactS})
is still exponentially bounded on the physical sheet. There are no poles either on the physical or on the unphysical sheets, however, there is a cut all the way to $s=\infty$, exhibiting an infinite number of broad resonances---excited string states (or, better to say, black holes, see \cite{critical}). 

A direct  indication that a ``free" critical string theory considered as a relativistic two-dimensional field theory is quite unusual, comes from its finite temperature properties. For a relativistic field theory the finite temperature free energy $f(T)$ is related to the finite volume ground state energy $E_0(R)$ as 
\[
f(T)= TE_0(T^{-1})\;.
\] 
By making use of the exact light-cone expression for $E_0$, we find that the free energy becomes complex above a critical temperature $T_H=(8\pi)^{-1/2}\ell^{-1}_s$.
Both the heat capacity and its temperature integral  diverge as one approaches this temperature, indicating that it is impossible to heat up the theory above $T_H$ with a finite amount of energy. Of course, these properties are very familiar to string theorists; this is the famous Hagedorn behavior of string theory. However, it is quite unusual when considered as a property of a renormalizable (in fact integrable) relativistic two-dimensional theory. In particular, this strongly suggests that unlike other known solvable theories with massless S-matrices, this model does not correspond to the RG flow between UV and IR conformal field theories. At low energies the theory  does flow
into a theory of 24 free bosons. However, its behavior in the UV is very unlikely to be described by a conventional fixed point, providing a new type of RG flow exhibited in a UV complete theory.

Note, that an integrable UV complete relativistic theory with S-matrix (\ref{exactS}) exists for any $D$, this is the integrable light cone theory ${\cal L}_{LC}$ of section~\ref{mass_spectrum}.
The special property of $D=26$ theory is that it exhibits a non-linearly realized Poincar\'e symmetry $ISO(D-1,1)$.
We feel that a detailed study of these models from the view-point adopted here deserves a separate publication, and we will present it in  \cite{critical}.

\section*{Acknowledgements}
We thank Ofer Aharony, Nima Arkani-Hamed, Giga Gabadadze, Walter Goldberger, Simeon Hellerman, Zohar Komargodski, Mehrdad Mirbabayi, Alberto Nicolis, Joe Polchinski, Massimo Porrati, Matt Roberts, Raman Sundrum, Peter Tinyakov, Arkady Tseytlin and Giovanni Villadoro for useful discussions and feedback.
This work is supported in part by the NSF grant PHY-1068438.
\bibliographystyle{utphys}
\bibliography{dlrrefs}

\providecommand{\href}[2]{#2}\begingroup\raggedright\begin{thebibliography}{10}

\bibitem{Teper:2009uf}
M.~Teper, ``{Large N and confining flux tubes as strings - a view from the
  lattice},'' {\em Acta Phys.Polon.} {\bf B40} (2009) 3249--3320,
  \href{http://www.arXiv.org/abs/0912.3339}{{\tt 0912.3339}}.

\bibitem{Kuti:2005xg}
J.~Kuti, ``{Lattice QCD and string theory},'' {\em PoS} {\bf LAT2005} (2006)
  001, \href{http://www.arXiv.org/abs/hep-lat/0511023}{{\tt hep-lat/0511023}}.

\bibitem{Low:2001bw}
I.~Low and A.~V. Manohar, ``{Spontaneously broken space-time symmetries and
  Goldstone's theorem},'' {\em Phys.Rev.Lett.} {\bf 88} (2002) 101602,
  \href{http://www.arXiv.org/abs/hep-th/0110285}{{\tt hep-th/0110285}}.

\bibitem{Coleman:1969sm}
S.~R. Coleman, J.~Wess, and B.~Zumino, ``{Structure of phenomenological
  Lagrangians. 1.},'' {\em Phys.Rev.} {\bf 177} (1969) 2239--2247.

\bibitem{Callan:1969sn}
C.~G.~J. Callan, S.~R. Coleman, J.~Wess, and B.~Zumino, ``{Structure of
  phenomenological Lagrangians. 2.},'' {\em Phys.Rev.} {\bf 177} (1969)
  2247--2250.

\bibitem{Isham:1971dv}
C.~Isham, A.~Salam, and J.~Strathdee, ``{Nonlinear realizations of space-time
  symmetries. Scalar and tensor gravity},'' {\em Annals Phys.} {\bf 62} (1971)
  98--119.

\bibitem{Volkov:1973vd}
D.~V. Volkov, ``{Phenomenological Lagrangians},'' {\em
  Fiz.Elem.Chast.Atom.Yadra} {\bf 4} (1973) 3--41.

\bibitem{Polyakov:1986cs}
A.~M. Polyakov, ``{Fine Structure of Strings},'' {\em Nucl.Phys.} {\bf B268}
  (1986) 406--412.

\bibitem{Kleinert:1986bk}
H.~Kleinert, ``{The Membrane Properties of Condensing Strings},'' {\em
  Phys.Lett.} {\bf B174} (1986) 335--338.

\bibitem{Aharony:2010db}
O.~Aharony and N.~Klinghoffer, ``{Corrections to Nambu-Goto energy levels from
  the effective string action},'' {\em JHEP} {\bf 1012} (2010) 058,
  \href{http://www.arXiv.org/abs/1008.2648}{{\tt 1008.2648}}.

\bibitem{Aharony:2010cx}
O.~Aharony and M.~Field, ``{On the effective theory of long open strings},''
  {\em JHEP} {\bf 1101} (2011) 065,
  \href{http://www.arXiv.org/abs/1008.2636}{{\tt 1008.2636}}.

\bibitem{Billo:2012da}
M.~Billo, M.~Caselle, F.~Gliozzi, M.~Meineri, and R.~Pellegrini, ``{The
  Lorentz-invariant boundary action of the confining string and its universal
  contribution to the inter-quark potential},'' {\em JHEP} {\bf 1205} (2012)
  130,
\href{http://www.arXiv.org/abs/1202.1984}{{\tt 1202.1984}}.

\bibitem{Georgi:1994qn}
H.~Georgi, ``{Effective field theory},'' {\em Ann.Rev.Nucl.Part.Sci.} {\bf 43}
  (1993) 209--252.

\bibitem{Polchinski:1991ax}
J.~Polchinski and A.~Strominger, ``{Effective string theory},'' {\em
  Phys.Rev.Lett.} {\bf 67} (1991) 1681--1684.

\bibitem{Polchinski:1992vg}
J.~Polchinski, ``{Strings and QCD?},''
\href{http://www.arXiv.org/abs/hep-th/9210045}{{\tt hep-th/9210045}}.

\bibitem{Polyakov:1981rd}
A.~M. Polyakov, ``{Quantum Geometry of Bosonic Strings},'' {\em Phys.Lett.}
  {\bf B103} (1981) 207--210.

\bibitem{Natsuume:1992ky}
M.~Natsuume, ``{Nonlinear sigma model for string solitons},'' {\em Phys.Rev.}
  {\bf D48} (1993) 835--838,
\href{http://www.arXiv.org/abs/hep-th/9206062}{{\tt hep-th/9206062}}.

\bibitem{Luscher:1980ac}
M.~Luscher, ``{Symmetry Breaking Aspects of the Roughening Transition in Gauge
  Theories},'' {\em Nucl.Phys.} {\bf B180} (1981) 317.

\bibitem{Luscher:2004ib}
M.~Luscher and P.~Weisz, ``{String excitation energies in SU(N) gauge theories
  beyond the free-string approximation},'' {\em JHEP} {\bf 0407} (2004) 014,
  \href{http://www.arXiv.org/abs/hep-th/0406205}{{\tt hep-th/0406205}}.

\bibitem{Aharony:2009gg}
O.~Aharony and E.~Karzbrun, ``{On the effective action of confining strings},''
  {\em JHEP} {\bf 0906} (2009) 012,
  \href{http://www.arXiv.org/abs/0903.1927}{{\tt 0903.1927}}.

\bibitem{Aharony:2011gb}
O.~Aharony and M.~Dodelson, ``{Effective String Theory and Nonlinear Lorentz
  Invariance},'' {\em JHEP} {\bf 1202} (2012) 008,
\href{http://www.arXiv.org/abs/1111.5758}{{\tt 1111.5758}}.

\bibitem{Gerstein:1971fm}
I.~Gerstein, R.~Jackiw, S.~Weinberg, and B.~Lee, ``{Chiral loops},'' {\em
  Phys.Rev.} {\bf D3} (1971) 2486--2492.

\bibitem{Cai:1994um}
W.~Cai, T.~Lubensky, P.~C. Nelson, and T.~Powers, ``{Measure factors, tension,
  and correlations of fluid membranes},'' {\em J. Phys. France II} {\bf 4}
  (1994) 931,
\href{http://www.arXiv.org/abs/cond-mat/9401020}{{\tt cond-mat/9401020}}.

\bibitem{critical}
S.~Dubovsky, R.~Flauger, and V.~Gorbenko, ``{Solving the Simplest Theory of
  Quantum Gravity},''
\href{http://www.arXiv.org/abs/1205.6805}{{\tt 1205.6805}}.

\bibitem{Polchinski:1998rq}
J.~Polchinski, ``{String theory. Vol. 1: An introduction to the bosonic
  string},''.

\bibitem{Adams:2006sv}
A.~Adams, N.~Arkani-Hamed, S.~Dubovsky, A.~Nicolis, and R.~Rattazzi,
  ``{Causality, analyticity and an IR obstruction to UV completion},'' {\em
  JHEP} {\bf 0610} (2006) 014,
  \href{http://www.arXiv.org/abs/hep-th/0602178}{{\tt hep-th/0602178}}.

\bibitem{Aharony:2011ga}
O.~Aharony, M.~Field, and N.~Klinghoffer, ``{The effective string spectrum in
  the orthogonal gauge},''
\href{http://www.arXiv.org/abs/1111.5757}{{\tt 1111.5757}}.

\bibitem{Dugan:1990df}
M.~J. Dugan and B.~Grinstein, ``{On the vanishing of evanescent operators},''
  {\em Phys.Lett.} {\bf B256} (1991)
239--244.

\bibitem{Cheng:2002iz}
H.-C. Cheng, K.~T. Matchev, and M.~Schmaltz, ``{Radiative corrections to
  Kaluza-Klein masses},'' {\em Phys.Rev.} {\bf D66} (2002) 036005,
\href{http://www.arXiv.org/abs/hep-ph/0204342}{{\tt hep-ph/0204342}}.

\bibitem{Luscher:1986pf}
M.~Luscher, ``{Volume Dependence of the Energy Spectrum in Massive Quantum
  Field Theories. 2. Scattering States},'' {\em Commun.Math.Phys.} {\bf 105}
  (1986)
153--188.

\bibitem{Luscher:1990ck}
M.~Luscher and U.~Wolff, ``{How To Calculate The Elastic Scattering Matrix In
  Two-Dimensional Quantum Field Theories By Numerical Simulation},'' {\em
  Nucl.Phys.} {\bf B339} (1990)
222--252.

\bibitem{Zamolodchikov:1991vx}
A.~Zamolodchikov, ``{From tricritical Ising to critical Ising by thermodynamic
  Bethe ansatz},'' {\em Nucl.Phys.} {\bf B358} (1991)
524--546.

\end{thebibliography}\endgroup
\end{document}